\documentclass[twocolumn,prl,amsmath,amssymb,showpacs,floatfix,superscriptaddress,citeautoscript,nofootinbib]{revtex4-1}

\usepackage{latexsym,epsfig,graphicx,subfigure}
\usepackage{dcolumn,verbatim}
\usepackage{bm,mathrsfs,bbold}
\usepackage{color}
\usepackage[colorlinks,urlcolor=blue,citecolor=blue]{hyperref}

\begin{document}

\title{General framework for transport in spin-orbit-coupled superconducting heterostructures: Nonuniform spin-orbit coupling and spin-orbit-active interfaces}

\author{Kuei Sun}
\affiliation{Department of Physics, University of Cincinnati,
Cincinnati, Ohio 45221-0011, USA} \affiliation{Department of
Physics, The University of Texas at Dallas, Richardson, Texas
75080-3021, USA}
\author{Nayana Shah}
\affiliation{Department of Physics, University of Cincinnati, Cincinnati,
Ohio 45221-0011, USA}

\pacs{74.45.+c, 74.25.F-, 71.70.Ej, 74.25.fc}

\begin{abstract}
Electronic spin-orbit coupling (SOC) is essential for various
newly discovered phenomena in condensed-matter systems. In
particular, one-dimensional topological heterostructures with SOC
have been widely investigated in both theory and experiment for
their distinct transport signatures indicating the presence of
emergent Majorana fermions. However, a general framework for the
SOC-affected transport in superconducting heterostructures,
especially with the consideration of interfacial effects, has not
been developed even regardless of the topological aspects. We
hereby provide one for an effectively one-dimensional
superconductor-normal heterostructure with nonuniform magnitude
and direction of both Rashba and Dresselhaus SOC as well as a
spin-orbit-active interface. We extend the
Blonder-Tinkham-Klapwijk treatment to analyze the current-voltage
relation and obtain a rich range of transport behaviors. Our work
provides a basis for characterizing fundamental physics arising
from spin-orbit interactions in heterostructures and its
implications for topological systems, spintronic applications, and
a whole variety of experimental setups.
\end{abstract}

\maketitle

\section{I.~Introduction}
The interplay between electronic spin and orbital
degrees of freedom, or spin-orbit coupling (SOC), has played a
crucial role in various aspects of condensed-matter physics,
including the study on semiconductors, ferromagnets,
superconductors and materials with exotic
orders~\cite{Kato04,Kane05,Bernevig06,Konig07,Wang09,Fischer08,Muhlbauer09,Lee09,Hamann11,Kruger12,Nadj-Perge10,Kloeffel11,Nadj-Perge12,Shanavas14}
as well as the application on building quantum electronic or
spintronic devices~\cite{Zutic04,Zwanenburg13}. Due to the
ubiquity of SOC and superconductivity, it is of fundamental and
technological interest to understand their interplay, in
particular, in the context of superconducting heterostructures.
Furthermore, great interest has been stimulated recently in
exploring topological states of matter~\cite{Hasan10,Qi11}, whose
experimental realization and detection strongly rely on transport
or scanning tunneling microscopy (STM) measurements of
artificially engineered heterostructures with SOC. Among them,
those with building blocks such as a
semiconductor/topological-insulator wire with SOC and conventional
superconductor~\cite{Fu08,Lutchyn10,Oreg10,Cook11,Mourik12} (used
to induce a proximity gap) or one-dimensional (1D) ferromagnet
combined with a bulk superconductors with
SOC~\cite{Duckheim11,Chung11,Nadj-Perge13,Nadj-Perge14,Hui14} are
of particular interest to explore emerging Majorana
fermions~\cite{Kitaev01,Bolech07,Wilczek09,Franz10,Stoudenmire11,Alicea12,Beenakker13}.
Most theoretical studies have focused on either effective models
or employed SOC as a uniform model
parameter~\cite{Law09,Qu11,Deng12,Liu12,Rokhinson12,Das12,Lin12,Roy12,Finck13,Churchill13,Jacquod13,Wu14b}.
However, the set-up of interest usually has only a segment or end
point of the 1D element in contact with a (bulk) superconductor
due to either the experimental constraints or an explicit interest
in studying (topological)
heterostructures~\cite{Mourik12,Wang10,Rokhinson12,Abay14}.

\begin{figure}[t]
\centering
  \includegraphics[width=8.6cm]{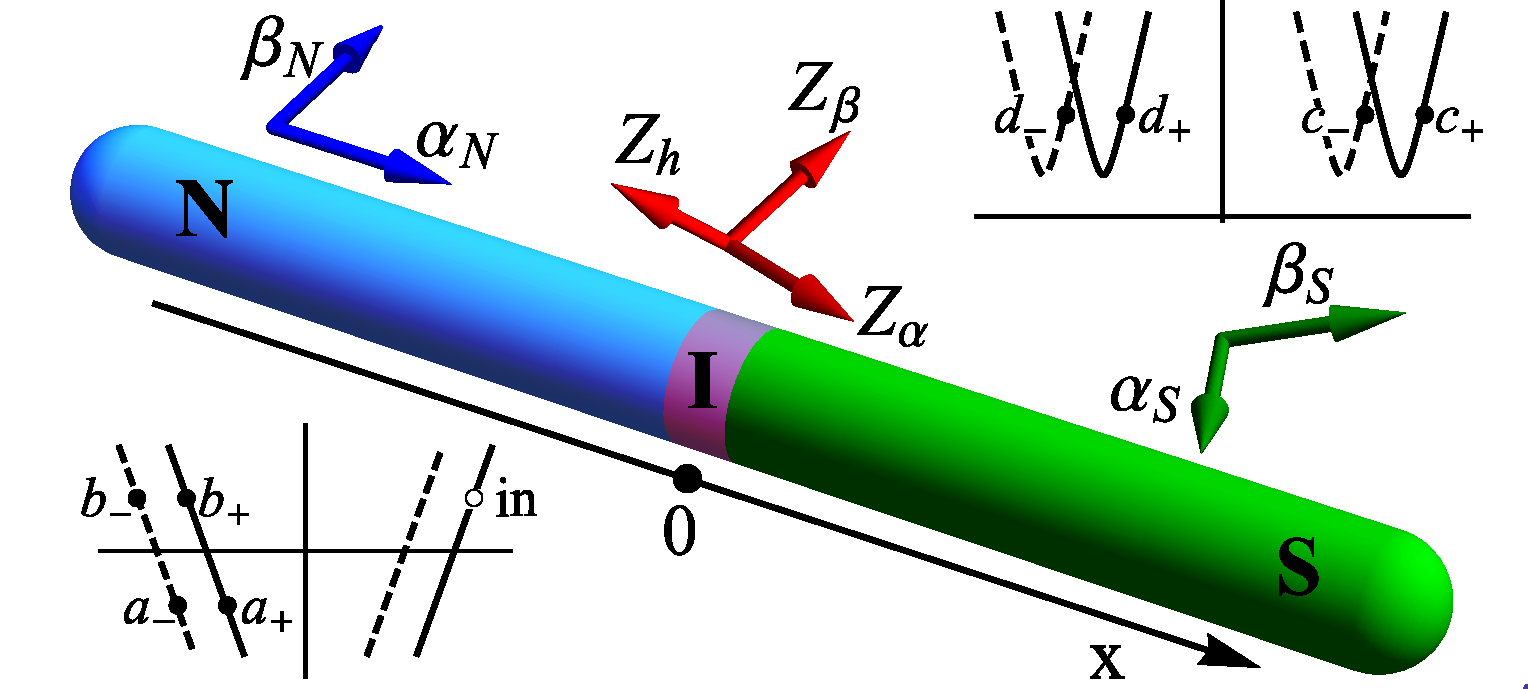}
  \vspace{-0.3cm}
  \caption{(Color online) Illustration of an effective 1D N--S system with SOC and a spin-orbit-active interface (I). The vectors denote
directions of Rashba $\alpha_{\rm{N(S)}} \sigma_y$ and Dresselhaus
$\beta_{\rm{N(S)}} \sigma_x$ on the N (S) side, as well as
interfacial Rashba $Z_\alpha \sigma_y$, Dresselhaus $Z_\alpha
\sigma_x$, and Zeeman $Z_h {\hat n} {\cdot} {\vec \sigma}$
components. In general, the spin basis in each region can be
different. Insets: schematic energy spectra with two isospin
branches (solid and dashed curves, respectively) on the
corresponding sides. The solid circles denote possible outgoing
states corresponding to an incoming electron (empty circle) from
the N side.} \label{fig:f1} \vspace{-0.4cm}
\end{figure}

In this paper, we develop a framework to study an (effectively) 1D
normal- (N-) superconductor (S) setup (see Fig.~\ref{fig:f1}) that
is general enough to capture the effects of SOC in a wide range of
(emergent) heterostructures.  We allow for both
Rashba~\cite{Rashba60} and Dresselhaus~\cite{Dresselhaus55} SOC
with arbitrary relative strengths on each side as well as at the
interface (I), which represents the boundary
region~\cite{interface}. Furthermore, we allow the SOC to have a
different magnitude as well as direction in N, S, and I regions.
Nonuniform SOC is in fact bound to arise for a variety of reasons
including difference in the chemical composition, crystallographic
direction, or direction of electric
field~\cite{Rashba60,Dresselhaus55} and may be relevant also for
practical implementations of topological quantum computing schemes
~\cite{Ojanen13,Klinovaja14}. SOC in the interface can even simply
arise due to a lack of \lq \lq left-right"
symmetry~\cite{Gorkov01,Edelstein03,Linder11}.

We start by analyzing the appropriate Bogoliubov--de Gennes (BdG)
Hamiltonian and develop a generalized Blonder-Tinkham-Klapwijk
(BTK)~\cite{Blonder82} formalism to calculate the current-voltage
characteristics for our general setup. We phenomenologically model
what we call the \emph{spin-orbit-active} interface as a
generalization of a \emph{spin-active}
interface~\cite{Buzdin05,Bergeret05,Xia02,Zhao04,Cottet05,Wu14a,Shevtsov14,Millis88,Tokuyasu88,Fogelstrom00,Ryazanov01,deJong95,Kalenkov07,Hubler12,Colci12,Sun13},
which itself is a generalization of a single $Z$ parameter used in
the original BTK approach.  We find that the rich physics
emanating from non-uniform SOC and/or spin-orbit-active interface
is not only of fundamental interest but also highly relevent for
interpreting the results of experiments. For example, it can
effectively lower the interfacial barrier, drastically change the
zero-bias conductance, and determine between reentrant or
monotonic temperature dependence. Our results also underline the
danger in using a single $Z$-parameter fit to extract the
interface properties in experiments as is standard practice. Our
analysis offers a more accurate picture even for traditional N-S
heterostructures or for STM setups in which a spin-orbit-active
interface or the nonzero SOC in the materials may play a role that
is not usually taken into account (although the importance of
nonzero SOC in the context of pair-breaking
effects~\cite{Maki69,Meservey94,Lopatin05,Rogachev05,Shah07,Maestro08}
is appreciated).

\section{II.~Model and Hamiltonian}

As illustrated in Fig.~\ref{fig:f1}, our effectively 1D system
comprises N ($x<0$) and S ($x>0$) regions with an arbitrary SOC on
each side, intersecting at I ($x=0$). The Hamiltonian
$H_{j=\rm{N,S}}$ in corresponding basis ${\Psi _j} = {\left(
{\begin{array}{*{20}{c}} {{\psi _{{ \uparrow _j}}}}&{{\psi _{{
\downarrow _j}}}}&{\psi _{{ \downarrow _j}}^\dag }&{ - \psi _{{
\uparrow _j}}^\dag }
\end{array}} \right)^T}$ reads
\begin{eqnarray}
{H_{\rm{N}}} &=& {\tau _z} \otimes \left( {{E_p}\mathbb{1} + {\alpha _{\rm{N}}}p{\sigma _y} + {\beta _{\rm{N}}}p{\sigma _x}} \right),\\
{H_{\rm{S}}} &=& {\tau _z} \otimes \left( {{E_p}\mathbb{1} +
{\alpha _{\rm{S}}}p{\sigma _y} + {\beta _{\rm{S}}}p{\sigma _x}}
\right) - \Delta {\tau _x} \otimes \mathbb{1},
\end{eqnarray}
where $E_p=\frac{p^2}{2m}-\mu$ is the free spectrum with chemical
potential $\mu$, $\alpha _j$ ($\beta_j$) are Rashba (Dresselhaus)
SOC on the $j$ side, $\Delta$ (taken real for convenience) is the
$s$-wave (possibly proximity-induced) superconducting gap, ${\vec
\sigma }$ and ${\vec \tau }$ are Pauli matrices spanned in spin
and particle-hole basis, respectively, and $\mathbb{1}$ is the $2
\times 2$ identity matrix. The spin basis on both sides can differ
by an Euler rotation
\begin{eqnarray}
{\Psi _{\rm{N}}} &=& \left( {\mathbb{1} \otimes {e^{ - i{\sigma
_z}\eta _{{\rm{NS}}}^{(3)}}}{e^{ - i{\sigma _y}\eta
_{{\rm{NS}}}^{(2)}}}{e^{ - i{\sigma _z}\eta _{{\rm{NS}}}^{(1)}}}}
\right){\Psi _{\rm{S}}} \nonumber\\
&\equiv& {U_{{\rm{NS}}}}{\Psi _{\rm{S}}}.
\end{eqnarray}
The two SOCs can be combined as a complex factor
\begin{eqnarray}
{\gamma _j}{e^{i{\varphi _j}}} = {\beta _j} + i{\alpha _j}.
\end{eqnarray}
We rotate the Hamiltonians into an isospin basis ${\Psi _{j'}}$
under transformation ${H_{j'}} = U_{jj'}^\dag {H_j}{U_{jj'}}$
where
\begin{eqnarray}
{U_{jj'}} = \frac{1}{{\sqrt 2 }}\mathbb{1} \otimes \left(
{\begin{array}{*{20}{c}}
{{e^{ - i{\varphi _j}}}}&1\\
{ - 1}&{{e^{i{\varphi _j}}}}
\end{array}} \right)
\end{eqnarray}
and obtain a diagonal ${H_{{\rm{N'}}}}$ and block-diagonal
${H_{{\rm{S'}}}}$ as
\begin{eqnarray}
{H_{{\rm{N'}}}} &=& {\tau _z} \otimes \left( {{E_p}\mathbb{1} -
{\gamma
_{\rm{N}}}p{\sigma _z}} \right),\\
{H_{{\rm{S'}}}} &=& {\tau _z} \otimes \left( {{E_p}\mathbb{1} -
{\gamma _{\rm{S}}}p{\sigma _z}} \right) - \Delta {\tau _x} \otimes
\mathbb{1}.
\end{eqnarray}
The spectra of ${H_{{\rm{N'}}}}$ and ${H_{{\rm{S'}}}}$ show
particle and hole bands, denoted by $\tau=\pm$. Each band has two
isospin branches denoted by $\sigma=\pm$. At a given energy $E$,
there are eight corresponding eigen wave functions for
${H_{{\rm{N'}}}}$ and the other eight for ${H_{{\rm{S'}}}}$,
denoted as
\begin{eqnarray}
\begin{array}{*{20}{c}} {\chi _{{\rm{N'}}}^{\tau ,\sigma }{e^{i( \pm p_\sigma ^\tau  +
\sigma {\gamma _{\rm{N}}})x}}\rm{~~and~}}&{\chi _{{\rm{S'}}}^{\tau
,\sigma }{e^{i( \pm k_\sigma ^\tau + \sigma {\gamma
_{\rm{S}}})x}},}
\end{array}
\end{eqnarray}
respectively, where
\begin{eqnarray}
p_\sigma ^\tau  &=& \sqrt {2m\left( {\mu ' - m\delta {\gamma ^2} +
\tau E} \right)},  \\
k_\sigma ^\tau &=& \sqrt {2m\left( {\mu ' + m\delta {\gamma ^2} +
\tau \sqrt {{E^2} - {\Delta ^2}} } \right)},
\end{eqnarray}
with
\begin{eqnarray}
\begin{array}{*{20}{c}} {\mu ' = \mu + \frac{m(\gamma _{\rm{N}}^2 + \gamma _{\rm{S}}^2)}{4} \rm{~~and~}}&{\delta {\gamma ^2} = \frac{\gamma _{\rm{S}}^2 - \gamma
_{\rm{N}}^2}{4}}.
\end{array}
\end{eqnarray}
The particle-hole band and isospin states have representations as
$\chi _{{\rm{N'}}}^{\tau ,\sigma } = {\left(
{\begin{array}{*{20}{c}} {{\delta _{\tau  + }}{\delta _{\sigma  +
}}}&{{\delta _{\tau  + }}{\delta _{\sigma  - }}}&{{\delta _{\tau -
}}{\delta _{\sigma  + }}}&{{\delta _{\tau  - }}{\delta _{\sigma -
}}}
\end{array}} \right)^T}$ and $\chi _{{\rm{S'}}}^{\tau ,\sigma } =
u{\left( {\begin{array}{*{20}{c}} {{\delta _{\tau  + }}{\delta
_{\sigma  + }}}&{{\delta _{\tau  + }}{\delta _{\sigma  -
}}}&{{\delta _{\tau  - }}{\delta _{\sigma  + }}}&{{\delta _{\tau
- }}{\delta _{\sigma  - }}}
\end{array}} \right)^T} + v{\left( {\begin{array}{*{20}{c}}
{{\delta _{\tau  - }}{\delta _{\sigma  + }}}&{{\delta _{\tau  -
}}{\delta _{\sigma  - }}}&{{\delta _{\tau  + }}{\delta _{\sigma  +
}}}&{{\delta _{\tau  + }}{\delta _{\sigma  - }}}
\end{array}} \right)^T}$, where ${u^2} = 1 - {v^2} =
\frac{1}{2}\left( {1 + \frac{{\sqrt {{E^2} - {\Delta ^2}} }}{E}}
\right)$ and $\delta$ is the delta function. Below we drop $\hbar$
in all equations for convenience and take $\mu '$ and $\sqrt {2 m
\mu '}$ to be natural energy and momentum units, respectively (so
$\gamma _{\rm{S}}^2 + \gamma _{\rm{N}}^2$ causes no qualitative
change but energy rescaling). Note that we consider energy range in which the
hole excitations have real momenta~\cite{ill-picture}.

In a BTK treatment, the interface is phenomenologically modeled by
Hamiltonian ${H_{\rm{I}}}\delta (x)$, which generally has the same
$4 \times 4$ representation as the bulk Hamiltonian. The matrix
components reflect the interfacial properties as well as the
physical discontinuity between both sides and specify transport
processes through the interface. The original BTK model adopts one
parameter $Z_0 \tau_z \otimes \mathbb{1}$ to describe barrier
effects from an oxide layer or local disorder~\cite{Blonder82}.
For ferromagnet--superconductor heterostructures, ${H_{\rm{I}}}$
can be modeled as a Zeeman form $\mathbb{1} \otimes Z_h \left(
{\hat n \cdot \vec \sigma } \right)$ with unit direction $\hat n =
\left( {\sin \theta \cos \phi ,\sin \theta \sin \phi ,\cos \theta
} \right)$, which accounts for various spin-active processes such
as spin-flip scattering, spin-dependent phase shift, and
spin-related Andreev
reflection~\cite{Xia02,Zhao04,Cottet05,Wu14a,Shevtsov14,Millis88,Tokuyasu88,Fogelstrom00,Ryazanov01,deJong95,Kalenkov07,Hubler12,Colci12,Sun13}.
In our case, we expect that the discontinuity of the bulk SOC and
the interfacial SOC itself would play an active role in the
transport. This effect is modeled by Rashba $Z_\alpha \tau_z
\otimes \sigma_y$ and Dresselhaus $Z_\beta \tau_z \otimes
\sigma_x$ components. To capture the most general interplay with
spins, we incorporate all the factors above and write down our
spin-orbit-active interface,
\begin{eqnarray}
{H_{\rm{I}}} = {\tau _z} \otimes \left( {{Z_0}\mathbb{1} +
{Z_\alpha }{\sigma _y} + {Z_\beta }{\sigma _x}} \right) +
\mathbb{1} \otimes {Z_h}\left( {\hat n \cdot \vec \sigma }
\right). \label{eq:HI}
\end{eqnarray}
The two SOCs can be combined as ${Z_\beta } + i{Z_\alpha } =
{Z_\gamma }{e^{i{\varphi _{\rm{I}}}}}$. In general, the spin basis
of ${H_{\rm{I}}}$ and ${H_{\rm{N}}}$ can differ by another Euler
rotation
\begin{eqnarray}
{U_{{\rm{NI}}}}=\left( {\mathbb{1} \otimes {e^{ - i{\sigma _z}\eta
_{{\rm{NI}}}^{(3)}}}{e^{ - i{\sigma _y}\eta
_{{\rm{NI}}}^{(2)}}}{e^{ - i{\sigma _z}\eta _{{\rm{NI}}}^{(1)}}}}
\right).
\end{eqnarray}
We write the rotated ${H_{\rm{I}}}$ in a suggestive form as
\begin{eqnarray}
&&{U_{{\rm{NI}}}}{H_{\rm{I}}}U_{{\rm{NI}}}^\dag  \nonumber\\
&& \equiv \left( {\begin{array}{*{20}{c}}
{{Z_0} - {Z_1}}&{{Z_2}{e^{ - i\zeta }}}&0&0\\
{{Z_2}{e^{i\zeta }}}&{{Z_0} + {Z_1}}&0&0\\
0&0&{ - {Z_0} - {Z_3}}&{ - {Z_4}{e^{ - i(\zeta  + \zeta ')}}}\\
0&0&{ - {Z_4}{e^{i(\zeta  + \zeta ')}}}&{ - {Z_0} + {Z_3}}
\end{array}} \right). \nonumber\\ \label{eq:HI2}
\end{eqnarray}
There are relations $Z_1^2 + Z_2^2 + Z_3^2 + Z_4^2 = Z_h^2 +
Z_\gamma ^2$ and $Z_1^2 + Z_2^2 - Z_3^2 - Z_4^2 = 2{Z_\gamma
}{Z_h}\sin \theta \cos \left( {\phi  - {\varphi _{\rm{I}}}}
\right)$ that are independent of the Euler angles. Comparing
Eq.~(\ref{eq:HI}) and Eq.~(\ref{eq:HI2}), one can see that the
three Euler angles effectively generate one additional variable.

In brief, the key ingredients on our system, the bulk SOCs
$\alpha_j \sigma_y$ and $\beta_j \sigma_x$ as well as the
interfacial parameters $Z_\alpha \sigma_y$, $Z_\beta \sigma_x$,
and $Z_h {\hat n} {\cdot} {\vec \sigma}$, can be characterized by
different vectors in spin space in the corresponding regions, as
illustrated in Fig.~\ref{fig:f1}. Below we study the system's
transport properties as a function of these vectors.

\section{III.~BTK calculations}

Here we apply the BTK treatment to compute current-voltage
relations of the system. Considering an incoming wave
\begin{eqnarray}
\Psi _{{\rm{in}}}^{\tau ,\sigma } = \chi _{{\rm{N'}}}^{\tau
,\sigma }{e^{i(\tau p_\sigma ^\tau x  + \sigma {\gamma
_{\rm{N}}})}},
\end{eqnarray}
with energy $E$ on the N side that propagates toward the interface
(positive group velocity) and scatters through, we incorporate all
possible outgoing waves and write down the wave functions on the N
and S sides, as
\begin{eqnarray}
{\Psi _{\rm{L}}}&=& {U_{{\rm{NN'}}}}\Big [ \Psi _{{\rm{in}}}^{\tau
,\sigma } + \sum\nolimits_{\sigma} e^{i \sigma \gamma_{\rm{N}}} \nonumber\\
&& \times \Big ( {b_\sigma }\chi _{{\rm{N'}}}^{ + ,\sigma }{e^{- i
p_\sigma ^ + x }} + {a_\sigma }\chi _{{\rm{N'}}}^{ - ,\sigma
}{e^{ip_\sigma ^ - x
}} \Big ) \Big ],\\
{\Psi _{\rm{R}}}&=&
{U_{{\rm{NS}}}}{U_{{\rm{SS'}}}}\sum\nolimits_{\sigma} e^{i \sigma
\gamma_{\rm{S}}} \nonumber\\
&& \times \Big ( {c_\sigma }\chi _{{\rm{S'}}}^{ + ,\sigma }{e^{i
k_\sigma ^ + x}} + {d_\sigma }\chi _{{\rm{S'}}}^{ - ,\sigma
}{e^{-i k_\sigma ^ - x }} \Big ),
\end{eqnarray}
respectively (all wave functions are represented in the real spin
basis of $H_{\rm{N}}$). Because reflections (transmissions) should
have the group velocity direction opposite to (same as) the
incoming wave, only eight states are considered (see insets in
Fig.~\ref{fig:f1}). For an incoming electron ($\tau=+$), the
amplitudes $b$, $a$, $c$, and $d$ correspond to normal reflection,
Andreev reflection~\cite{Andreev64}, quasi-particle transmission
and quasi-hole transmission, respectively, while for an incoming
hole ($-$), $a$ and $b$ reverse their roles. The subscript
$\sigma$ of the amplitudes describes in-branch (cross-branch)
processes if its sign is the same as (opposite to) the incoming
wave. These amplitudes can be determined by two boundary
conditions at the interface. The first one is the continuity of
the wavefunction
\begin{eqnarray}
{\Psi _{\rm{L}}}(0)={\Psi _{\rm{R}}}(0),
\end{eqnarray}
while the second one can be obtained by integrating the BdG
equation over an infinitesimal interval across the interface,
\begin{eqnarray}
\int_{{0^ - }}^{{0^ + }} \Big \{ && \frac{1}{2} \big [
{H_{\rm{N}}}\theta ( - x) + {U_{{\rm{NS}}}}{H_{\rm{S}}}
U_{{\rm{NS}}}^\dag \theta (x) + \rm{H.c.} \big ] \nonumber\\ && +
{U_{{\rm{NI}}}}{H_{\rm{I}}}U_{{\rm{NI}}}^\dag \delta (x) \Big \}
\Psi dx = 0,
\end{eqnarray}
where $\theta(x)$ is the step function and $\Psi ({0^ \pm }) =
{\Psi _{\rm{R/L}}}(0)$. We carefully keep the Hamiltonian
Hermitian when it is expressed using the step
function~\cite{hermitian}.

The probability current $J$ is calculated from its definition
${\partial _t}{\Psi ^\dag }\Psi  =  - {\partial _x}J$. For each
incoming wave $\Psi _{{\rm{in}}}^{\tau ,\sigma }$, we calculate
$J$ corresponding to different scattering processes normalized by
the incoming current and obtain combined currents carried by the
in- and cross-branch normal (Andreev) reflections together,
$J_{\tau ,\sigma }^{{\rm{NR}}}$ ($J_{\tau ,\sigma }^{{\rm{AR}}}$),
as well as those carried by all the transmissions together,
$J_{\tau ,\sigma }^{{\rm{T}}}$ ($= 1-J_{\tau ,\sigma
}^{{\rm{NR}}}-J_{\tau ,\sigma }^{{\rm{AR}}}$ due to the
probability conservation). Following the standard BTK treatment,
the net charge current $I$ induced by a voltage drop $V$ across
the junction can be evaluated as
\begin{eqnarray}
I &=& \sum\nolimits_{\tau ,\sigma} Ae\tau \int_0^\infty  {dE \Big
\{ \left( {1 - J_{\tau ,\sigma }^{{\rm{NR}}}} \right)  \Big [ {f(E
- \tau eV) - f(E)} \Big ]} \nonumber\\ && + J_{\tau ,\sigma
}^{{\rm{AR}}} \Big [ {f(E) - f(E + \tau eV)}  \Big ] \Big \},
\end{eqnarray}
where $A$ is a constant associated with density of states, Fermi
velocity, as well as an effective cross-sectional area, and
$f(E)=[\rm{exp}(E/k_BT)+1]^{-1}$ is the Fermi distribution
function at temperature $T$. We compute the $I$-$V$ relation and
normalized differential conductance (NDC) $G/G_0$, where
$G=\frac{dI}{dV}$ and $G_0=\frac{dI(\Delta=0)}{dV}$ is a reference
value when the S side is normal.

\section{IV.~Results}

We first find that $\sum\nolimits_{\sigma } {J_{\tau ,\sigma
}^{{\rm{NR/AR/T}}}}$ and hence $I(V)$ are independent of the Euler
angles $\eta _{{\rm{NS}}}^{(i)}$. Therefore, parameters such as
$\varphi_{\rm{N}}$ and $\varphi_{\rm{S}}$ that rely on the
relative spin coordinate (RSC) between both sides should play no
role on the transport either. We numerically confirm this
independence by varying the ratio of Rashba and Dresselhaus SOC on
each side. For the interface, $I(V)$ is independent of $\eta
_{{\rm{NI}}}^{(i)}$ if one of $Z_h$ and $Z_\gamma$ in
Eq.~(\ref{eq:HI}) is zero, so $\hat n$ or $\varphi_{\rm{I}}$ has
no effect given $Z_{\gamma}=0$ or $Z_{h}=0$, respectively. We find
that the scattering amplitude and probability current for each
channel sensitively vary with RSC, but they compensate in the
summation for $I(V)$. This implies that RSC may affect $I(V)$ as a
result of interference in a multichannel or multiterminal
system~\cite{Colci12,Sun13} and also open interesting
possibilities for spintronics.

\begin{figure}[t]
\centering
\includegraphics[width=8.6cm]{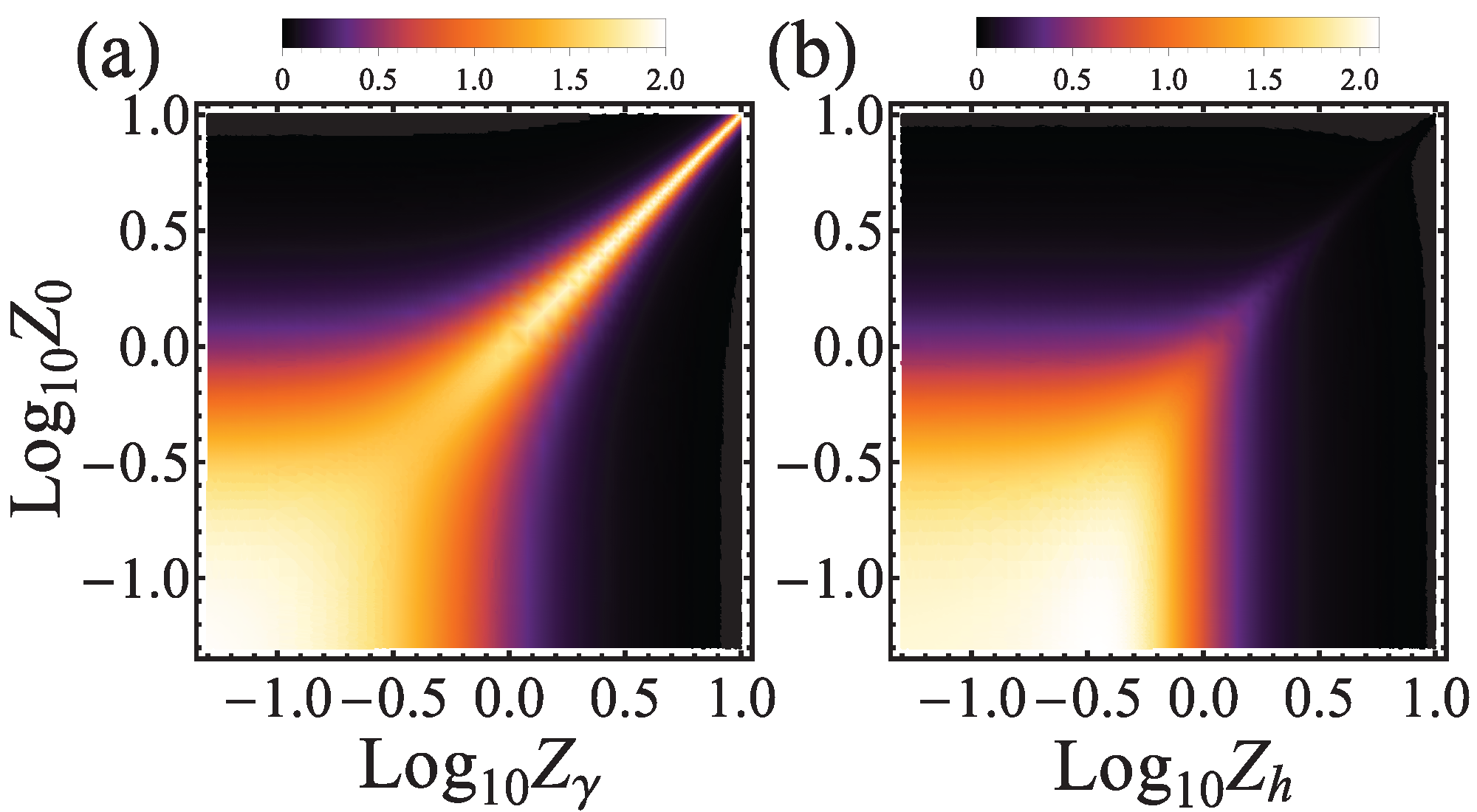}
\vspace{-0.3cm}
  \caption{(Color online) (a) [(b)] Zero-bias NDC (scaled by colors in bar graph) in the
$Z_0$-$Z_\gamma$ ($Z_h$) plane given $Z_h~(Z_\gamma)=0$. Data are
for $\Delta=0.002$, $\delta \gamma^2=0$, and $T=0$.}
\label{fig:f2}
\end{figure}

In Fig.~\ref{fig:f2}(a) we plot zero-bias NDC in the
$Z_0$-$Z_\gamma$ plane for a purely spin-orbit-active interface
($Z_h=0$). We see that NDC reaches a high value when $Z_\gamma$ is
close to $Z_0$ and exhibits a symmetry under the exchange between
$Z_0$ and $Z_\gamma$. To explain this, one can Euler rotate
$H_{\rm{I}}$ to its eigenspin basis. The eigenvalues can be
regarded as a set of characteristic barrier strengths $\left|Z_0
\pm Z_\gamma \right|$ that determine the BTK results. Such set is
invariant under $Z_0 \leftrightarrow Z_\gamma$ and hence leads to
the symmetry.  To understand the NDC peaks, one can look at the
scattering of the eigenspin states. At $Z_\gamma=Z_0$, electrons
of one spin direction and holes of the opposite both see a clean
interface $\left|Z_0 - Z_\gamma \right|=0$, which maximizes the
Andreev current as well as NDC. In other words, the presence of
interfacial SOC can effectively lower the original BTK barrier
($\left|Z_0 - Z_\gamma \right|<Z_0$)~\cite{non-monotonic}. In
Fig.~\ref{fig:f2}(b) we plot NDC in the $Z_0$-$Z_h$ plane for a
purely spin-active interface $Z_\gamma=0$ for comparison. We see
that there is no symmetry under $Z_0 \leftrightarrow Z_h$ and no
conductance peak along the $Z_0 = Z_h$ line. Such differences
illustrate the interfacial SOC effects that the original
spin-active picture does not capture.

\begin{figure}[t]
\centering
\includegraphics[width=8.6cm]{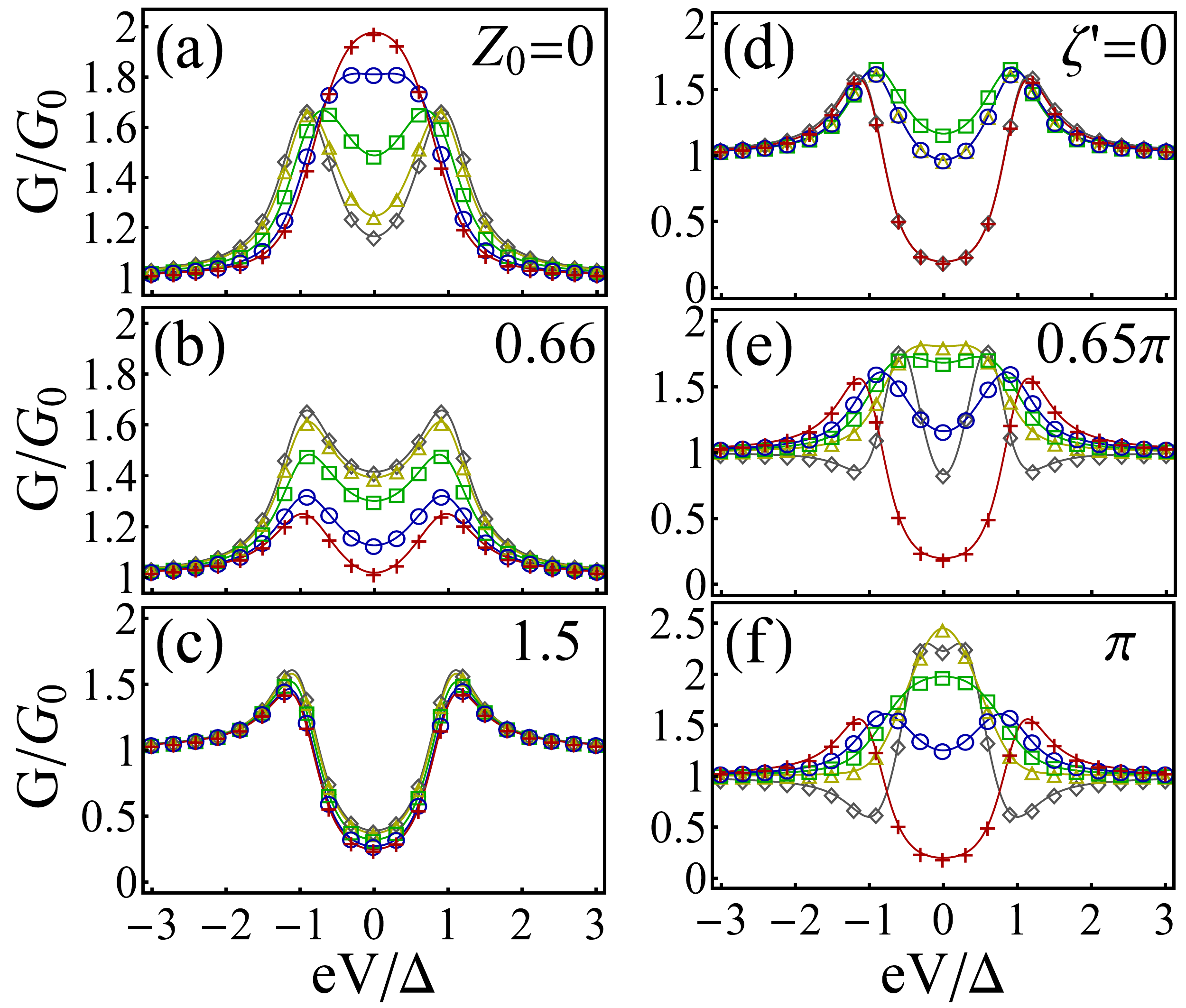}
\vspace{-0.3cm}
  \caption{(Color online) (a)--(c) Interfacial phase effects on NDC for
  a general spin-orbit-active interface described by Eq.~(\ref{eq:HI2}) with $Z_0=0$, $0.66$, and
  $1.5$, respectively.
  Curves with diamonds, triangles, squares, circles, and crosses
  correspond to $\zeta'=0$, $0.25\pi$, $0.5\pi$, $0.75\pi$, and
  $\pi$, respectively. The other relevant parameters are set as
  $Z_1=Z_3=0$, $Z_2=Z_4=0.5$, $\Delta=0.002$, $\delta \gamma^2=0$,
  and $T=0.15 \Delta$. (d)--(f) Charge imbalance effects on NDC at
  cases of $\zeta'=0$, $0.65\pi$, and $\pi$, respectively.
  Curves with diamonds, triangles, squares, circles, and crosses
  correspond to $m\delta \gamma^2=-0.975$, $-0.5$, $0$, $0.5$, and
  $0.975$, respectively. The other parameters are the same as
  (a)--(c) except $Z_0=0$ and $\delta \gamma^2$ varies.
  }
\label{fig:f3}
\end{figure}

With the coexistence of Zeeman and SOC effects at the interface,
the Euler angles $\eta _{{\rm{NI}}}^{(i)}$ are no longer
irrelevant. We consider the general Hamiltonian of
Eq.~(\ref{eq:HI2}) and find that $I(V)$ is independent of $\zeta$.
The relevant variables are the five strength parameters
$Z_{0,1,2,3,4}$ and the phase difference $\zeta '$ between
off-diagonal elements of the particle and hole blocks. Notice that
the role of $\zeta '$ is special: (1) it comes from the interplay
between $Z_h$ and $Z_\gamma$ and (2) it does not alter the
eigenvalues of the interfacial Hamiltonian. Therefore, its effects
on the transport can be attributed to the interference between
particle and hole channels. In
Figs.~\ref{fig:f3}(a)--\ref{fig:f3}(c), we plot NDC vs $V$ at
various $\zeta '$ and $Z_0$ (we set $Z_1=Z_3=0$ and $Z_2=Z_4=0.5$
for illustrating the salient features from the interplay between
$\zeta'$ and $Z_0$). At $Z_0=0$ (a), the curves are all above 1
and show a qualitative change from center-dent to center-peak
types as $\zeta '$ goes through $0$ (diamonds), $0.25\pi$
(triangles), $0.5\pi$ (squares), $0.75\pi$ (circles), and $\pi$
(crosses). In this half period, the zero-bias NDC monotonically
increases with $\zeta '$. From $\zeta ' = \pi$ to $\zeta ' =
2\pi$, the deformation of curves completes the other half period
and reverses back to the $\zeta ' = 0$ case. As $Z_0$ increases,
the center-peak curves drastically deform toward the center-dent
type, and the trend of zero-bias NDC vs $\zeta '$ also reverses.
At $Z_0=0.66$ (b), the $\zeta ' = \pi$ curve has the lowest
zero-bias NDC $\approx 1$. At $Z_0=1.5$ (c), the center-dent
curves remain and are well below 1 at low bias due to the
suppression of transmissions and Andreev reflections (tunneling
limit in the BTK model). The role of $\zeta '$ is not significant
in this case. We turn to show one way to independently tune $\zeta
'$ via the tuning of parameters in Eq.~(\ref{eq:HI}). Assuming the
same spin basis on the N side and the interface ($\eta
_{{\rm{NI}}}^{(i)}=0$) and the interfacial Zeeman components tuned
as $\theta=\pi/2$, $Z_h \cos \phi = Z_\gamma \sin^2
\varphi_{\rm{I}}/\cos \varphi_{\rm{I}}$, and $Z_h \sin \phi =
-Z_\gamma \sin \varphi_{\rm{I}}$, we obtain $Z_1=Z_3=0$,
$Z_2=Z_4=Z_\gamma/\cos \varphi_{\rm{I}}$, and $\zeta '=-2
\varphi_{\rm{I}}$ by equating Eqs.~(\ref{eq:HI}) and
(\ref{eq:HI2}). In this case, $\zeta '$ is associated with the
ratio of Rashba and Dresselhaus components at the interface with
properly controlled Zeeman components.

All the results above are for cases of balanced charge carriers
between N and S sides ($\gamma_{\rm{N}}=\gamma_{\rm{S}}$). In
Figs.~\ref{fig:f3}(d)--\ref{fig:f3}(f) we explore the effects of
nonuniform SOC induced imbalance ($\delta \gamma ^2 \neq 0$
resulting in Fermi momentum mismatch) and its interplay with the
interfacial parameters. At $\zeta ' =0$, the curves are all of the
center-dent type and roughly display symmetry between positive and
negative imbalance. The zero-bias NDC is far below 1 in the highly
imbalanced cases. As $\zeta '$ increases, the $\delta \gamma^2 \le
0$ curves deform more drastically than the $\delta \gamma^2 > 0$
ones. At $\zeta ' =0.65 \pi$, the $m \delta \gamma^2=-0.5$ curve
develops a plateau around $V=0$, indicating an incipience of a
center peak. The $m \delta \gamma^2=-0.975$ curve develops
minimums around $eV=\Delta$. At $\zeta '=\pi$, the $m \delta
\gamma^2=-0.5$ and $0$ curves show a center-peak structure, and
the double minimums of the $m \delta \gamma^2=-0.975$ curves
become more significant. The range of zero-bias NDC as a function
of $\delta \gamma^2$ also maximizes. These rich behaviors can be
attributed to the high mismatch between the quasi-particle momenta
on both sides and the interfacial phase difference $\zeta '$
between particle and hole channels, which together alter the
scattering amplitudes in the BTK calculations.

\begin{figure}[t]
\centering
\includegraphics[width=8.6cm]{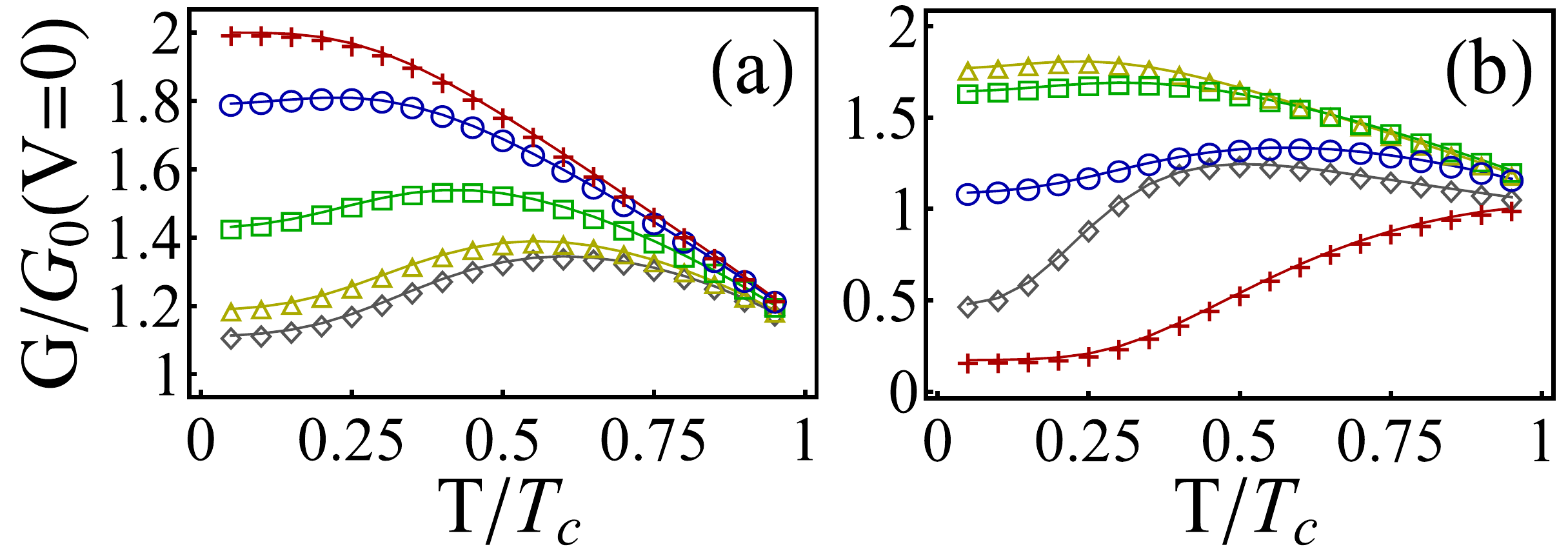}
\vspace{-0.3cm}
  \caption{(Color online) (a) Zero-bias NDC vs $T$ at various $\zeta '$
  [convention and setting are the same as Fig.~\ref{fig:f3}(a)].
  (b) Zero-bias NDC vs $T$ at various $\delta \gamma ^2$ [same as Fig.~\ref{fig:f3}(e)].
  }
\label{fig:f4}
\end{figure}

Finally we discuss the temperature dependence of the transport. In
Fig.~\ref{fig:f4} we plot zero-bias NDC vs $T/T_c$ ($T_c$ is the
superconducting transition temperature) at various $\zeta '$ [(a),
same convention and setting as Fig.~\ref{fig:f3}(a)] or $\delta
\gamma^2$ [(b), same as Fig.~\ref{fig:f3}(e)]. The curves can
exhibit three types of behavior: (1) monotonic increase, (2)
monotonic decrease, and (3) first increase and then decrease (a
reentrant phenomenon). These rich behaviors come from the same
reason as the case of NDC vs $V$ do, because high-energy
excitations play a more important role at higher temperature. As a
result, the conductance as a function of temperature is also
sensitive to the bulk SOC and spin-orbit-active interface.

\section{V.~Conclusion}

In conclusion, a general framework for superconducting
heterostructures with SOC is developed and analyzed by
generalizing the BTK scheme. Study of the conductance reveals the
crucial role spin-orbit-active interface and nonuniform SOC play
in determining the transport properties. In addition to being
directly relevant for underpinning the physics of a range of
hybrid systems of relevance to various fields including
topological matter and spintronics, our work opens up many future
research avenues. One is to apply and extend our transport
analysis to other heterostructures, for example, to study the
interplay with magnetic fields or ferromagnetism and also to study
the multichannel interference effects in quasi-one-dimensional,
higher-dimensional or multiterminal
setups~\cite{Shah06,Colci12,Sun13}, also taking into account
effects of fluctuations when a low-dimensional
superconductor~\cite{Sahu09,*Sau11,*Fidkowski11,*Brenner11,*Brenner12}
is involved. Another is to analyze the impact on pairing symmetry,
such as the possibility of $p$-wave
pairing~\cite{Volkov03,*Bolech04,*Keizer06,*Eschrig08,*Linder09,*Almog11,*Leksin12,*Tanaka12,*Bergeret13,*Gentile13,*Sun14a,*Liu15}
and its consequences, and to establish a general framework also
using complementary approaches such as quasiclassic
formalism~\cite{Demler97,*Stanev14,*Alidoust15}. Finally, another
important direction is to explore the consequences of
nonuniformity and boundary/interface behaviors of SOC in ultracold
atoms~\cite{Lin11,*Liu12b,*Wang12,*Cheuk12,*Qu13,*Zhang13,*Chen13,*Sun14b,*Qu15}
by leveraging their tunability and amenability for probing the
dynamics.

\textbf{Acknowledgements}: This work is supported by University of
Cincinnati.


\begin{thebibliography}{}
\bibitem{interface}The interface could be a thin insulating layer,
a layer with disorder, or simply a boundary between N and S
regions (which could be a narrow region with rapidly changing
physical properties).
\bibitem{ill-picture}In the high imbalance limit $\delta {\gamma ^2} \to \pm 1$, the hole excitations have imaginary momenta in a wide range of $E$ and thus become ill-defined in the BdG scenario.
\bibitem{hermitian}For example, the discontinuity of SOC between N and S sides should
not be described by linear momentum terms simply as $p \theta (x)$
but a Hermitian form $\frac{1}{2} [p \theta (x)+\rm{h.c.}]$.
\bibitem{non-monotonic}Note that along the line $Z_0 = Z_\gamma$, the minimum of NDC appears at a middle point due to a non-monotonic relative behavior between $G$ and $G_0$.
\end{thebibliography}
\end{document}